%% file: main.tex
\documentclass{article}

\usepackage{PRIMEarxiv}

\usepackage[utf8]{inputenc} % allow utf-8 input
\usepackage[T1]{fontenc}    % use 8-bit T1 fonts
\usepackage{hyperref}       % hyperlinks
\usepackage{url}            % simple URL typesetting
\usepackage{booktabs}       % professional-quality tables
\usepackage{amsfonts}       % blackboard math symbols
\usepackage{nicefrac}       % compact symbols for 1/2, etc.
\usepackage{microtype}      % microtypography
\usepackage{lipsum}
\usepackage{fancyhdr}       % header
\usepackage{graphicx}       % graphics
\usepackage{amsmath}
\usepackage{float}
\usepackage{graphicx}
\usepackage{amssymb}
\usepackage[normalem]{ulem}
\usepackage{tabularray}
\usepackage{subcaption}
\usepackage[table,xcdraw]{xcolor}
\usepackage{pifont}
\usepackage[normalem]{ulem}
\usepackage{multirow} 
\useunder{\uline}{\ul}{}
\title{MLLA-UNet: Mamba-like Linear Attention in an Efficient U-Shape Model for Medical Image Segmentation
\thanks{\textit{\underline{Citation}}: 
\textbf{Authors. Title. Pages.... DOI:000000/11111.}} 
}

\author{
    Yufeng Jiang\hspace{.3mm}\textsuperscript{\rm 1},
    Zongxi Li\hspace{.3mm}\textsuperscript{\rm 2}\thanks{Corresponding author} ,
    Xiangyan Chen\hspace{.3mm}\textsuperscript{\rm 1},
    Haoran Xie\hspace{.3mm}\textsuperscript{\rm 2},
    Jing Cai\hspace{.3mm}\textsuperscript{\rm 1}$^\dagger$,
    \\     $^1$ Department of Health Technology and Informatics, The Hong Kong Polytechnic University, Hong Kong SAR \\
     $^2$ School of Data Science, Lingnan University, Hong Kong SAR\\
  \texttt{yufeng.jiang@connect.polyu.hk}\\
  \texttt{zongxili@ln.edu.hk}\\
  \texttt{jing.cai@polyu.edu.hk}\\
 }

\begin{document}
\maketitle

\begin{abstract}
Recent advancements in medical imaging have resulted in more complex and diverse images, with challenges such as high anatomical variability, blurred tissue boundaries, low organ contrast, and noise.
Traditional segmentation methods struggle to address these challenges, making deep learning approaches, particularly U-shaped architectures, increasingly prominent. 
However, the quadratic complexity of standard self-attention makes Transformers computationally prohibitive for high-resolution images. 
% v3
To address these challenges, we propose MLLA-UNet (\textit{Mamba-Like Linear Attention UNet}), a novel architecture that achieves linear computational complexity while maintaining high segmentation accuracy through its innovative combination of linear attention and Mamba-inspired adaptive mechanisms, complemented by an efficient symmetric sampling structure for enhanced feature processing.
Our architecture effectively preserves essential spatial features while capturing long-range dependencies at reduced computational complexity. 
Additionally, we introduce a novel sampling strategy for multi-scale feature fusion. 
Experiments demonstrate that MLLA-UNet achieves state-of-the-art performance on six challenging datasets with 24 different segmentation tasks, including but not limited to FLARE22, AMOS CT, and ACDC, with an average DSC of 88.32\%. 
These results underscore the superiority of MLLA-UNet over existing methods. Our contributions include the novel 2D segmentation architecture and its empirical validation.
The code is available via \href{https://github.com/csyfjiang/MLLA-UNet}{this link}.
\end{abstract}

% keywords can be removed
\keywords{2D Medical Image Segmentation\and Semantic Segmentation \and Linear Attention \and UNet \and Vision SSM}

\input{sec/0_introducition}
\input{sec/1_Related_Work}
\input{sec/2_Methods}

\input{sec/3_Experiments}

\input{sec/4_Reults}
\input{sec/5_Discussion}
\input{sec/6_Conclusion}

%Bibliography
\bibliographystyle{plain}  
\bibliography{references}

\end{document}

%% file: sec/0_introducition.tex
\section{Introduction}
% ---
Medical image segmentation is critical in computer-aided diagnosis and treatment planning. 
With advancements in medical imaging technology, we are confronted with increasingly complex and diverse medical images. 
These images often exhibit high anatomical variability, blurred tissue boundaries, low contrast between organs, and imaging noise and artifacts \cite{metheany2008characterizing}. 
Traditional segmentation methods often struggle with these complexities, necessitating more advanced techniques to address these challenges effectively~\cite{ma2010review_tm,Litjens2017,vaswani2017attention_transformer}.
% ---

% ---
In recent years, deep learning approaches, particularly those based on U-shaped architectures, have gained prominence in medical image segmentation. 
Traditional U-shaped models like UNet and its variants~\cite{long2015fully_cnn,brosch2016deep_cnn,milletari2016v_cnn} utilize convolutional neural networks (CNNs)~\cite{lecun1998gradient_cnn}, which excel at capturing local features and hierarchical representations. 
However, CNNs with limited receptive fields cannot extract long-range dependencies that are essential for understanding the global context of anatomical structures~\cite{chen2021transunet_tran}. 
This limitation hinders medical segmentation especially when organs with large shape and size variations across patients.
To address this challenge, Transformer-based medical segmentation models with self-attention mechanism~\cite{vaswani2017attention_transformer} have emerged as a promising alternative~\cite{lin2022ds_tran,swinunet,swinunetr,chen2021transunet_tran,wu2024medsegdiff_tran_diff,vaswani2017attention_transformer,dosovitskiy2020image_vit}. 
However, they face challenges in preserving local structural information that is critical for accurate boundary delineation. 
Additionally, the quadratic computational complexity of standard self-attention with respect to input size makes Transformers computationally expensive for high-resolution medical images~\cite{swinunet}. 
Linear attention offers linear computational complexity, but it suffers from insufficient expressiveness compared to traditional attention mechanisms~\cite{katharopoulos2020transformers_linearAt}.
% ---
% ---
Recently, Mamba-based medical image segmentation models gained remarkable development, including U-Mamba~\cite{ma2024umamba}, VM-UNet~\cite{ruan2024vmunet,zhang2024vmunetv2}, Swin-UMamba~\cite{liu2024swinumamba}, \textit{inter alia}. 
%%% SSM
These State Space Models (SSMs) leverage the Mamba structure's advantages, such as efficient long-range dependency and spatial feature extraction, adaptation to various image sizes and multi-scale features, and efficient training on larger image patches. 
These characteristics make SSMs particularly suited for the complex medical image segmentation task, where understanding local details and global context are crucial.
However, these methods still require the recursive computation in Mamba's forget gate~\cite{han2024demystify_mlla}, which may not be suitable for non-autoregressive visual tasks, and potential deficiencies in preserving local detail information.
%%% v3
MLLA~\cite{han2024demystify_mlla} is essentially linear attention but with an improved design that approximates the selective SSM mechanism, combining the parallel processing efficiency of attention with the adaptive feature selection capability of SSMs. 
The linear attention mechanism enables efficient processing of high-resolution images with $O(n)$ complexity. At the same time, the SSM-inspired design provides selective feature focusing and noise filtering capabilities, making it particularly effective for medical image segmentation tasks where both computational efficiency and precise feature extraction are crucial.

% ---

% ---
To address the challenges faced by CNNs, Transformers, and Vision SSMs in medical image segmentation, we propose the \textbf{M}amba-\textbf{L}ike \textbf{L}inear \textbf{A}ttention \textbf{UNet} (MLLA-UNet). Our model builds upon the MLLA mechanism~\cite{han2024demystify_mlla}, which combines the advantages of linear attention and SSM frameworks.
%
% First explain MLLA mechanism
The MLLA mechanism integrates two key components: linear attention and Mamba-inspired selective mechanisms. The linear attention reduces computational complexity from $O(n^2)$ to $O(n)$ \cite{katharopoulos2020transformers_linearAt}, enabling efficient processing of high-resolution medical images. Meanwhile, the Mamba-inspired design \cite{visionMamba_203} provides adaptive feature selection capabilities, addressing the insufficient expressiveness of traditional linear attention while maintaining parallel computation advantages.
%
% Then introduce MLLA-UNet architecture advantages
Building on these foundational capabilities, our MLLA-UNet architecture adopts a symmetric U-shaped structure specifically designed for medical image segmentation. This design effectively leverages MLLA's advantages while addressing the specific challenges of medical image segmentation through efficient multi-scale feature processing.

% ---
% MLLA-UNet pros 3 from UNet % MLLA downsampling
Our architecture consists of three main components: a Stem module for initial feature extraction, feature compression stages for multi-scale representation learning, and feature expansion stages for precise reconstruction.
Han et al.~\cite{han2024demystify_mlla} originally devise MLLA for visual encoding, which consists of a stem module, MLLA blocks, and an \textbf{E}fficient \textbf{D}own\textbf{S}ampling \textbf{M}odule (EDSM).
More concretely, the stem module transforms the raw input into feature embeddings through a series of convolutions, capturing initial spatial information.
The feature compression stage combines MLLA blocks with EDSM to progressively reduce spatial dimensions while capturing long-range dependencies, effectively compressing the input into multi-scale representations.
% ---
To adapt to medical image segmentation tasks, we develop complementary feature expansion stages and skip connections.
Specifically, the feature expansion stage consists of MLLA blocks and our proposed \textbf{E}fficient \textbf{U}p\textbf{S}ampling \textbf{M}odule (EUSM), which incorporates depthwise-separable convolutions~\cite{szegedy2017inception} for upsampling operations.
This design allows for efficient parameter utilization while maintaining the ability to learn complex spatial relationships, potentially enhancing the model's capacity to adapt to diverse morphologies of different organs and tissues.
In contrast to simple linear operations used in previous methods~\cite{liu2024swinumamba,swinunet}, our approach is capable of capturing the intricate details necessary for accurate medical image segmentation.
% ---

% ---
% MLLA-UNet pros 4
We further explore MLLA-UNet's scalability by expanding the number of MLLA blocks at each layer and increasing the embedding dimensions. 
Our findings are consistent with the observations made by Gao et al. in their study \cite{gao2024training}, where they noted that in the traditional paradigm, the performance gains from increasing model scale are marginal due to the limited size and diversity of individual datasets. 
This contrasts with the neural scaling laws \cite{dosovitskiy2020image_vit,kaplan2020scaling,liu2021swin,he2016deep}, which suggest that larger models can lead to overfitting when the data is not sufficiently extensive. 
Our experimental results show that simultaneously scaling up both model size and dataset size led to substantial performance improvements, revealing the full potential of large models.
This synergistic approach corroborates the conclusions drawn by Gao et al. \cite{gao2024training} and establishes a practical pathway to achieve higher accuracy in medical image segmentation by maintaining an optimal balance between model complexity and data volume.
% comment1(done):
% ---

% ---
% MLLA-UNet overcome the vision ssm cons
% ---
MLLA-UNet's linear attention mechanism provides notable advantages over previous approaches. 
The combination of linear attention with Mamba-inspired designs allows parallel computation while maintaining expressiveness, making it particularly effective for medical image segmentation tasks. 
This design achieves both computational efficiency and high segmentation accuracy, especially for complex anatomical structures where precise boundary delineation is crucial.
% ---
% ---

% ---
% short results overview
We have conducted extensive experiments to test the effectiveness of our proposed MLLA-UNet on six challenging datasets with 24 different segmentation tasks.
Our proposed MLLA-UNet achieves state-of-the-art (SOTA) performance across multiple challenging medical image segmentation datasets, consistently outperforming existing models in both accuracy and efficiency. 
In particular, MLLA\textsubscript{Tiny} with only 34.14M parameters and 14.66G FLOPs attains an average Dice Similarity Coefficient (DSC) of 88.32\%, significantly exceeding the 86.34\% achieved by SwinUNetR~\cite{swinunetr}, a leading model in the field. 
Notably, our method reduces computational costs by 38.5\% compared to ConvNeXtv2 (23.82G FLOPs) while delivering superior performance. Additionally, our method demonstrates superior performance across key organ segmentation tasks, achieving a DSC of 89.1\% on the WORD dataset with an HD95 of 9.37 mm, effectively demonstrating its robustness and efficiency compared to other SOTA models.
% ---

% ---
% summarize the contribution
The main contributions of this paper can be summarized as follows:

\begin{itemize}
    \item We propose a novel medical image segmentation architecture based on Mamba-like linear attention, significantly enhancing segmentation accuracy and computational efficiency by combining the strengths of linear attention and SSM. 

    \item We implement a symmetric and efficient sampling strategy to preserve local structural information, by developing an upsampling method based on the original MLLA downsampling approach \cite{han2024demystify_mlla}. This unified design optimizes both feature extraction and reconstruction processes, enhancing the overall performance of our segmentation architecture.

    \item Our proposed MLLA-UNet achieves the SOTA performance across multiple challenging medical image datasets, including AMOS22CT/MR~\cite{ji2022amos}, WORD~\cite{luo2021word}, FLARE22~\cite{FLARE22}, ATLAS23~\cite{quinton2023tumour_altas}, BTCV~\cite{btcv}, and ACDC~\cite{acdc}. Our approach outperforms representative models from different architectural paradigms, including ConvNeXtv2 (CNN-based), Swin Transformer (Transformer-based), and VSS (Mamba-based), showcasing both the broad applicability and superiority of our method in diverse medical imaging contexts.
    
    \item We have conducted extensive experiments, providing in-depth insights into the advantages of MLLA in medical image segmentation. Additionally, we develop a scalable framework that paves the way for new research directions in future medical image analysis tasks, such as lesion detection and classification, real-time surgical navigation, dynamic organ tracking, and lightweight mobile deployment.

\end{itemize}
% ---

%% file: sec/1_Related_Work.tex
\section{Related Work}

\subsection{Visual Transformers and Self-Attention Mechanisms}

Vision Transformers (ViTs) have revolutionized computer vision by adapting the transformer architecture from natural language processing \cite{dosovitskiy2020image_vit}. The core of these models is the self-attention mechanism, which captures long-range dependencies in data.
Self-attention projects input features $X$ into queries $Q$, keys $K$, and values $V$: $Q = XW_q, K = XW_k, V = XW_v$. The attention distribution is computed as $A = \text{softmax} (\frac{QK^T}{\sqrt{d}})$, and the output as $Z = AV$. This enables global context modeling \cite{vaswani2017attention_transformer}.
ViTs process images by splitting them into patches, which are linearly projected and combined with positional embeddings. These are then processed through transformer blocks consisting of multi-head self-attention and feed-forward networks.

Standard self-attention has a quadratic complexity $O(n^2)$, which can be prohibitive for high-resolution images. Linear attention methods have been proposed to reduce this to $O(n)$ \cite{katharopoulos2020transformers_linearAt}.
Transformers excel at modeling long-range dependencies and preserving fine-grained details. They demonstrate improved robustness to image corruptions and favorable scaling properties. However, they lack the inductive biases of CNNs, potentially leading to increased data requirements and training difficulty.
The flexibility of transformer architectures has sparked numerous innovations, including hybrid models that combine the strengths of both CNNs and transformers \cite{cnn_vits_1,cnn_vits_2,cnn_vits_3,cnn_vits_4}.

\subsection{Mamba and Linear Attention in Vision}

Recent advancements in vision models have shifted towards more efficient architectures that maintain high performance while reducing computational complexity. This section reviews the emergence of Mamba-based models and Linear Attention mechanisms in vision tasks, highlighting their contributions to parameter efficiency and improved performance.

Introducing Mamba-based architectures has led to a new class of models capable of handling various vision tasks. As shown in Table 3, these models address various challenges in computer vision. Vision Mamba \cite{visionMamba_203} and VMamba \cite{liu2024vmamba_94} introduced bidirectional Mamba blocks and VSS blocks with SS2D modules, respectively, for classification, detection, and segmentation tasks. Mamba-ND \cite{li2024mamba-nd} extended the Mamba architecture to handle arbitrary multi-dimensional data, broadening its applicability to action recognition and forecasting. LocalMamba \cite{huang2024localmamba} and EfficientVMamba \cite{pei2024efficientvmamba} focused on improving local dependency capture and lightweight model design for visual tasks. More specialized models like SiMBA \cite{patro2024simba}, PlainMamba \cite{yang2024plainmamba}, and FractalVMamba \cite{tang2024scalable} introduced innovations such as Einstein FFT for channel modeling, non-hierarchical continuous 2D scanning, and fractal scanning curves for improved spatial relationships. These models demonstrate the versatility of Mamba-based architectures in addressing various aspects of visual understanding, from primary classification to complex segmentation tasks.

Parallel to Mamba developments, Linear Attention mechanisms have emerged as a solution to the quadratic complexity problem in traditional Vision Transformers. Parameter-Efficient Vision Transformer with Linear Attention \cite{zhao2023parameter} introduced the Linear Feature Attention (LFA) module, creating a hybrid CNN-ViT model called LightFormer. This model achieved competitive performance on ImageNet-1K with only 5.5 million parameters, demonstrating its efficiency in various visual recognition tasks. Mobile Attention \cite{yaomobile} addressed the efficiency-capability dilemma in mobile applications by proposing a head-competition mechanism. This approach enables linear-time complexity on mobile devices while maintaining model capability, making it suitable for resource-constrained environments. FLatten Transformer \cite{han2023flatten} introduced Focused Linear Attention to overcome limitations in current linear attention approaches. This model achieved improved performance by analyzing focus ability and feature diversity while maintaining low computational complexity. However, to our knowledge, no work has yet applied Linear Attention specifically to medical image segmentation tasks.

\subsection{Mamba Models for 2D Medical Image Segmentation}

Recent advancements in 2D medical image segmentation have seen a significant rise in Mamba-based architectures, each offering innovative solutions for various medical imaging modalities and anatomical structures.

U-Mamba~\cite{ma2024umamba} and Mamba-UNet~\cite{wang2024mambaunet} introduced hybrid CNN-SSM structures and symmetrical encoder-decoder architectures for multi-modal imaging (CT, MRI, etc.) and abdominal CT/MRI segmentation, respectively. The VM-UNet~\cite{ruan2024vmunet} series focused on abdominal and skin lesion segmentation, incorporating asymmetrical encoder-decoders and semantic detail infusion modules. Swin-UMamba~\cite{liu2024swinumamba} combined Transformer and Mamba architectures to enhance segmentation capabilities for endoscopic and microscopic MRI images. P-Mamba~\cite{ye2024pmamba} explored new approaches for pediatric echocardiography segmentation. ViM-UNet~\cite{archit2024vim} showcased the advantages of Vision Mamba architecture in microscopy image segmentation. SliceMamba~\cite{fan2024slicemamba} improved the segmentation accuracy for skin lesions and polyps through a bidirectional slice scan module.

These Mamba-series models demonstrate strong potential in 2D medical image segmentation tasks, covering various applications from CT and MRI to endoscopic and microscopic images. They provide more accurate and efficient tools for medical image analysis, pushing the boundaries of what is possible in this critical field.

%% file: sec/2_Methods.tex
\section{Methodology}

\begin{figure}[!htbp]
    \centering
    \includegraphics[width=.9\linewidth]{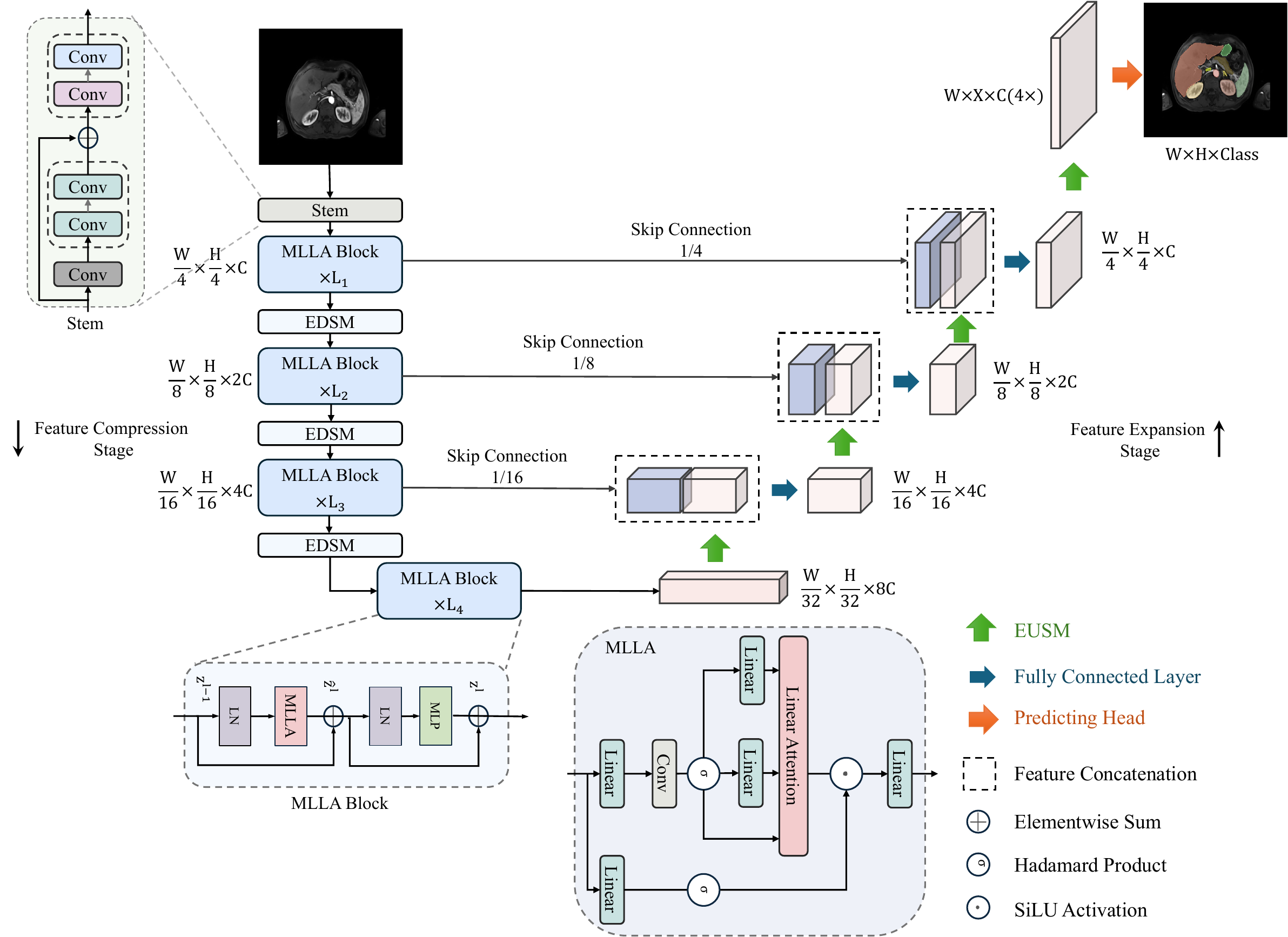}
    \caption{Architecture Overview: The model consists of a stem for initial processing, followed by multiple MLLA Blocks for feature extraction at various scales. EDSM layers reduce spatial dimensions, while skip connections preserve information across different layers. EUSM layers reconstruct the output, aided by fully connected layers for classification. Feature concatenation merges information from different paths, and final expanding adjusts the output to the desired dimensions and class count.}
    \label{fig:model}
\end{figure}

\subsection{MLLA-UNet architecture}
Our proposed MLLA-UNet adopts a U-shaped structure designed for medical image segmentation, with three key components: the stem module, feature compression stages, and feature expansion stages. This structure facilitates efficient multi-scale feature extraction and precise segmentation of complex anatomical structures by combining MLLA with a novel multi-scale fusion strategy. Figure~\ref{fig:model} provides an overview of the proposed model and its key components.

\subsubsection{Stem}
The stem module serves as the initial processing block following the design proposed in~\cite{han2024demystify_mlla}, efficiently converting the input image into feature embeddings. It performs patch-based embedding by gradually reducing spatial dimensions and increasing channel dimensions through convolutional operations. The stem module prepares the input for subsequent feature compression stages and efficiently maintains significant spatial information:
\begin{equation}
x_0 = f_{conv3}(f_{conv2}(f_{conv1}(x_{input})) + f_{conv1}(x_{input})),
\end{equation}
where $f_{conv1}$, $f_{conv2}$, and $f_{conv3}$ represent the sequential convolutional operations.

\subsubsection{Feature Compression}
The feature compression stages apply MLLA, which will be elaborated in Section \ref{sec:mlla_block}, to capture long-range dependencies while reducing the spatial resolution of the input. 
At each of the $L$ compression stages, MLLA blocks are employed to handle complex anatomical structures across various image sizes:
\begin{equation}
x_i = f_{EDSM,i}(f_{MLLA,i}(x_{i-1})),
\end{equation}
where $f_{MLLA,i}$ stands for the MLLA block at stage $i$, and $f_{EDSM,i}$ denotes the $i$-th down-sampling operation that progressively increases feature map channels.

\subsubsection{Feature Expansion}
A key innovation of our approach lies in the feature expansion stages, which mirror the compression process and aim to reconstruct the input image's spatial dimensions. At each stage, upsampling operations are combined with MLLA blocks to enhance the segmentation precision:
\begin{equation}
y_i = f_{MLLA,i}(f_{EUSM,i}(y_{i+1}) + x_{L-i}),
\end{equation}
where $f_{EUSM, i}$ represents the $i-th$ upsampling operation, $x_{L-i}$ denotes the skip connection from the corresponding compression stage, and $L$ represents the total number of layers in the MLLA-UNet network. Detailed feature compression and expansion operations will be discussed in more detail in Section~\ref{subsec:down_up_sampling}.

\subsection{Mamba-Like Linear Attention (MLLA) Block}
\label{sec:mlla_block}

The MLLA block~\cite{han2024demystify_mlla}, as shown in Figure~\ref{fig:model}, is a key component of our architecture, designed to efficiently capture long-range dependencies while maintaining linear complexity with $\mathcal{O}(N)$. The MLLA is defined as follows:
\begin{equation}
\begin{split}
F_1 &= \mathcal{L}(\text{Conv}(x)) \\
F_2 &= \mathcal{L}(x) \odot F_1 \\
F_3 &= \mathcal{L}(x) \\
\text{Attention} &= \text{LinearAttention}(F_2, F_3) \\
\text{Output} &= \mathcal{L}(\text{Attention})
\end{split}
\end{equation}
where $\mathcal{L}$ denotes linear transformation and $\odot$ represents the Hadamard (element-wise) product. The LinearAttention operation for position $i$ is defined as:
\begin{equation}
Q = \phi(x\mathbf{W}_Q), K = \phi(x\mathbf{W}_K), V = x\mathbf{W}_V,
\end{equation}
where $\phi$ is an additional kernel function, $x$ is the input, and $\mathbf{W}_Q$, $\mathbf{W}_K$, and $\mathbf{W}_V$ are learnable weight matrices for query, key, and value projections, respectively,
\begin{equation}
y_i = \frac{Q_i \left(\sum_{j=1}^N K_j^\top V_j\right)}{Q_i \left(\sum_{j=1}^N K_j^\top\right)},
\end{equation}
which can be reformulated into a more efficient recurrent form:
\begin{equation}
y_i = \frac{Q_i S_i}{Q_i Z_i}, \quad S_i = \sum_{j=1}^i K_j^\top V_j, \quad Z_i = \sum_{j=1}^i K_j^\top,
\end{equation}
resulting in a recurrent linear attention form:
\begin{equation}
S_i = S_{i-1} + K_i^\top V_i, \quad Z_i = Z_{i-1} + K_i^\top, \quad y_i = Q_i S_i / Q_i Z_i,
\end{equation}

\subsubsection{Position Encoding Strategy for Mamba-like Mechanism}

To address the limitation of recurrent computation in Mamba while maintaining its modeling capabilities, MLLA employs a combination of three position encoding techniques:

1) Locally-Enhanced Positional Encoding (LePE) Positional Encoding (LePE)~\cite{lepe}:
\begin{equation}
\text{LePE}(x) = x + \text{DWConv}(x)\mathbf{W}_L,
\end{equation}
where $\mathbf{W}_L$ is a learnable weight matrix, and DWConv denotes depth-wise convolution with kernel size $k$. LePE provides local bias similar to Mamba's forget gate.

2) Conditional Positional Encoding (CPE)~\cite{cpe}:
\begin{equation}
\text{CPE}(Q,K,V) = Q \cdot f_Q(p) + K \cdot f_K(p) + V \cdot f_V(p),
\end{equation}
where $p$ represents position indices, and $f_Q$, $f_K$, $f_V$ are learnable functions that map positions to encoding vectors.  offers input-dependent positional information.

3) Rotary Position Encoding (RoPE)~\cite{rope}:
\begin{equation}
\text{RoPE}(x_m, \theta_i) = x_m \cdot (\cos(m\theta_i) + \sin(m\theta_i)),
\end{equation}
where $x_m$ is the $m$-th dimension of the input, and $\theta_i$ is the position-dependent angle. RoPE provides global positional information.

MLLA replaces Mamba's recurrent forget gate with positional encodings to maintain parallel computation:
\begin{equation}
\text{PE}(x) = \text{CPE}_1(x) + \text{Attn}(\text{RoPE}(Q), \text{RoPE}(K), V + \text{LePE}(V)) + \text{CPE}_2(x)
\end{equation}
where $\text{CPE}_1$ and $\text{CPE}_2$ are convolutional positional encodings that capture hierarchical local spatial dependencies through depthwise convolutions, $\text{Attn}$ represents the linear attention operation that achieves $O(n)$ complexity through efficient query-key-value interactions, $\text{RoPE}$ applies rotary position embedding to queries and keys, encoding relative positional information in a way that preserves temporal order, $\text{LePE}$ adds local positional bias to values through depthwise convolution, enhancing the model's ability to capture fine-grained spatial patterns.

The MLLA block integrates these position encodings with MLP layers and layer normalization:
\begin{equation}
\begin{split}
\hat{z}^{l} &= \text{MLLA}(\text{LN}(z^{l-1})) + z^{l-1},\\
z^{l} &= \text{MLP}(\text{LN}(\hat{z}^{l})) + \hat{z}^{l},    
\end{split}
\end{equation}
where $\hat{z}^{l}$ and $z^{l}$ represent the output yielded from the MLLA module and MLP module of the $l^{th}$ block, respectively, and LN denotes Layer Normalization. This combination of position encodings enables MLLA to effectively capture local and global dependencies while maintaining parallel computation capability.

\subsection{EDSM and EUSM}
\label{subsec:down_up_sampling}
\begin{figure}
    \centering
    \includegraphics[width=0.8\linewidth]{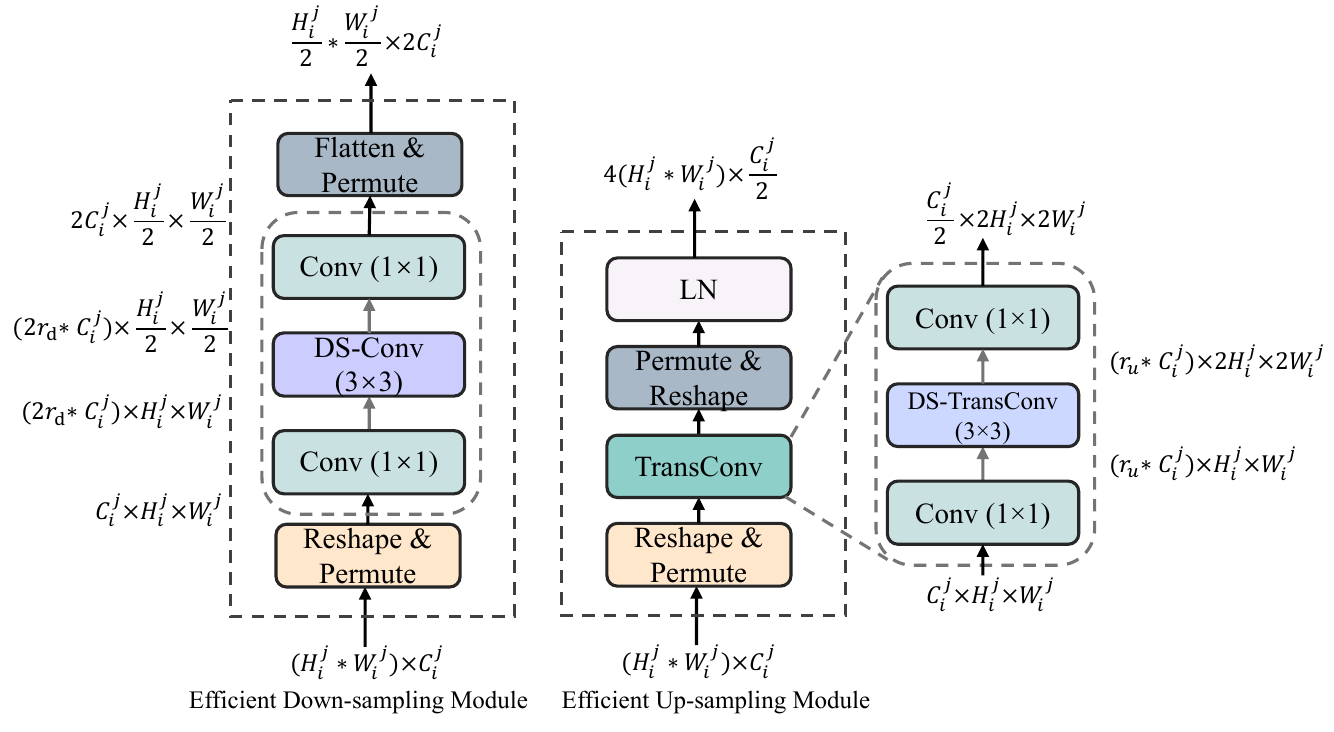}
    \caption{Illustration of EDSM and EUSM.}
    \label{fig:down-up}
\end{figure}
The EDSM~\cite{han2024demystify_mlla}, as shown in Figure~\ref{fig:down-up}, plays a pivotal role in reducing spatial dimensions while simultaneously increasing the number of channels, which can be formulated as:
\begin{equation}
\begin{split}
x_{out} &= f_{EDSM}(x_{in})\\
&=  f_{flatten}(f_{conv1\times1}(f_{DSconv3\times3}(f_{conv1\times1}(f_{reshape}(x_{in}))))),
\end{split}
\end{equation}
where $x_{in}$ with shape of $(H_i^j \times W_i^j) \times C_i^j$ and $x_{out}$ with shape of $\frac{H_i^j}{2} \times \frac{W_i^j}{2} \times 2C_i^j$. The number of channels at stage $i$ is progressively increased as:
\begin{equation}
C_i = C_0 \cdot 2^i,
\end{equation}
the inverse relationship between spatial resolution and channel depth allows for richer feature representations.

Conversely, the proposed EUSM, as depicted in Figure~\ref{fig:down-up}, is designed to increase spatial dimensions while reducing the number of channels to reconstruct the original image. This process can be expressed as:
\begin{equation}
\begin{split}
y_{out} &= f_{EUSM}(y_{in}) \\
&= LN(f_{reshape}(f_{conv1\times 1}(f_{TransConv3\times 3}(f_{conv1x\times 1}(f_{reshape}(y_{in})))))),
\end{split}
\end{equation}
where $y_{in}$ has shape $(H_i^j \times W_i^j) \times C_i^j$ and $y_{out}$ has shape $4\times(H_i^j \times W_i^j) \times \frac{C_i^j}{2}$. The number of channels decreases progressively according to the following:
\begin{equation}
D_i = D_L \cdot 2^{-(L-i)},
\end{equation}
where $D_i$ is the number of channels at decoder stage $i$, and $D_L$ is the number of channels at the bottleneck. The deep-wise separable transposed convolution (DS-TransConv3$\times$3) is key to doubling the spatial dimensions while the surrounding operations adjust channel counts and normalize the expanded features.
These carefully crafted down-sampling and up-sampling operations are fundamental to the network's ability to process and reconstruct features at multiple scales. 
These operations facilitate the capture of both fine-grained details and global context by enabling the model to navigate between different levels of spatial resolution and feature abstraction. 

\subsection{Predicting head}
Our predicting head transforms the feature maps to match the target segmentation dimensions:
\begin{equation}
\begin{split}
F &\in \mathbb{R}^{W \times H \times C(4\times)} \xrightarrow{\text{FinalPatchExpand}} \mathbb{R}^{4W \times 4H \times C} \xrightarrow{\text{Conv2d}} \mathbb{R}^{4W \times 4H \times C_{class}}
\end{split}
\end{equation}

where $F$ represents the input features, where $W$ and $H$ denote the spatial dimensions, and $C$ indicates the channel dimension. The transformation process involves first applying a spatial expansion operation that increases the resolution by a factor of 4 in both spatial dimensions while preserving the feature information. The transformation process is followed by a 1×1 convolutional layer that projects the feature space into pixel-level class-specific probability maps, where $C_{class}$ represents the number of segmentation classes.

\subsection{Loss function}
The segmentation loss combines Cross-Entropy and Dice loss:
\begin{equation}
\mathcal{L}_{total} = \alpha \mathcal{L}_{CE} + \beta \mathcal{L}_{Dice},
\end{equation}
where
\begin{equation}
\mathcal{L}_{CE} = -\sum_{c=1}^{C_{class}} y_c \log(\hat{y}_c),
\end{equation}

\begin{equation}
\mathcal{L}_{Dice} = 1 - \frac{2\sum_{c=1}^{C_{class}} y_c\hat{y}_c}{\sum_{c=1}^{C_{class}} y_c^2 + \sum_{c=1}^{C_{class}} \hat{y}_c^2},
\end{equation}
with $\alpha=0.4$ and $\beta=0.6$, $y_c$ as ground truth, and $\hat{y}_c$ as predicted probabilities for class $c$.

%% file: sec/3_Experiments.tex
\section{Experiments}

\subsection{Implementation details}

The experiments were conducted using PyTorch 2.2.0 as the deep learning framework. The models were trained on a system equipped with RTX 4090 24GB. We applied random transformations to the input images for data augmentation, including scaling and rotation. For training procedure. We employed the AdamW optimizer with a base learning rate of $0.0001$ and a weight decay of $0.01$. A Cosine Annealing Learning Rate Scheduler was used to adjust the learning rate over epochs, starting from the base learning rate and decaying to $1e-6$. The training and validation batch sizes were set according to the dataset specifics, with a typical batch size of 48. 

\subsection{Evaluation metrics}
% The performance of the models was evaluated using the Dice Similarity Coefficient (DSC) and Hausdorff Distance (HD95). These metrics provide insights into the overlap and boundary accuracy of the predicted segmentation. 
Our experiments use two primary metrics for evaluating segmentation performance: the Hausdorff Distance (HD95) and the Dice Similarity Coefficient (DSC).

The Hausdorff Distance (HD95) measures the 95th percentile of the maximum distances between two sets $X$ and $Y$, reducing sensitivity to outliers. It is defined as:
\begin{equation}
\begin{aligned}
    HD_{95}(X,Y) &= \max \{ h(X,Y), h(Y,X) \}, \\
    h(X,Y) &= \max_{x \in X} \min_{y \in Y} d(x,y), \\
    h(Y,X) &= \max_{y \in Y} \min_{x \in X} d(x,y),
\end{aligned}
\end{equation}
where $h(X, Y)$ and $h(Y, X)$ are the directed Hausdorff distances between $X$ (predicted points) and $Y$ (ground truth points), with $d(x,y)$ representing the distance between points $x$ and $y$.

We also use the Dice Similarity Coefficient (DSC) to evaluate the overlap between the predicted and ground truth segmentations:
\begin{equation}
    DSC(X,Y) = \frac{2 * |X \cap Y|}{|X| + |Y|},
\end{equation}
where $|X \cap Y|$ is the intersection size of sets $X$ and $Y$. DSC ranges from 0 to 1, with values closer to 1 indicating better segmentation overlap.

The best model was selected based on validation performance. We saved model checkpoints when an improvement in validation Dice Score was observed. The final model was saved after completing the predetermined number of iterations or epochs.
% comment1

\subsection{Datasets}
Our experiments leveraged six diverse medical image segmentation datasets: FLARE22~\cite{FLARE22} (13 abdominal organs, CT, 50 labeled cases), AMOS22~\cite{ji2022amos} (15 abdominal organs, CT and MRI, 500 CT and 100 MRI scans), ATLAS23~\cite{quinton2023tumour_altas} (liver and liver tumor, T1 CE-MRI, 60 training cases), WORD~\cite{luo2021word} (16 abdominal organs, CT, 150 scans), BTCV~\cite{btcv} (13 abdominal organs, CT, 50 cases), and ACDC~\cite{acdc} (cardiac structures, MRI, 150 examinations). For each dataset, we follow the ~\cite{huang2023stunet,swinunet,chen2021transunet_tran} for data split. To ensure consistency across all datasets, we strictly adhered to the nnUNet\cite{nnunet} standard pipeline for data preprocessing.
% comment1:

%% file: sec/4_Reults.tex
\section{Results}
\subsection{Results on multiple datasets of Medical Image Segmentation}

\definecolor{Black}{rgb}{0,0,0}
\begin{table}%[t]
\small
\centering
\caption{Comparison of segmentation methods on the WORD, FLARE22, AMOS CT, ALTAS23, BTCV, ACDC, and AMOS MR datasets. The `Avg' column represents the average DSC (\%). The best results are highlighted in \textbf{bold}, and the second-best results are in \uline{underlined}.}
\label{tab:overlall}
\begin{tblr}{
  width = \linewidth,
  colspec = {lcccccccc},
  % cells = {c},
  hline{1,2,5,12-13} = {-}{},
}
Methods & WORD & FLARE22 & AMOSCT & ALTAS23 & BTCV & ACDC & AMOSMR & Avg\\
Sun et al.~\cite{sunetal} & - & \uline{89.70} & - & - & - & - & - & -\\
Ma et al.~\cite{ma2024umamba} & - & - & 86.83 & 78.48 & 81.23 & 89.03 & 85.01 & -\\
MSVM-UNet~\cite{chen2024msvmunet} & - & - & - & - & 85.00 & \uline{92.58} & - & -\\
STU-Net-B~\cite{huang2023stunet} & 87.19 & 86.56 & \uline{89.84} & 79.01 & 83.29 & 90.18 & \uline{86.92} & 86.14\\
nnUNetV2~\cite{nnunetv2} & 85.8 & 88.37 & 86.53 & 80.22 & 80.17 & 89.36 & 77.19 & 83.95\\
nnUNetV1~\cite{nnunet} & 83.21 & 84.19 & 83.01 & 79.71 & 78.15 & 87.15 & 75.19 & 81.52\\
TransUNet~\cite{chen2021transunet_tran} & 79.12 & 83.5 & 79.30 & 76.53 & 76.76 & 89.71 & 75.12 & 80.01\\
SwinUNet~\cite{swinunet} & 82.14 & 84.61 & 82.53 & 76.84 & 79.13 & 90.00 & 78.12 & 81.91\\
SwinUNetR~\cite{swinunetr} & \uline{88.34} & 89.13 & 88.00 & \uline{79.15} & 84.53 & 91.87 & 83.35 & 86.34\\
UNetR~\cite{unetr} & 77.41 & 83.37 & 76.20 & 78.51 & \uline{84.73} & 83.24 & 60.38 & 77.69\\
\textbf{MLLA-UNet (Ours)} & \textbf{89.10} & \textbf{90.15} & \textbf{90.05} & \textbf{83.09} & \textbf{85.28} & \textbf{93.28} & \textbf{87.29} & \textbf{88.32}
\end{tblr}
\end{table}

\begin{figure}[!hb]
    \centering
    \includegraphics[width=1\linewidth]{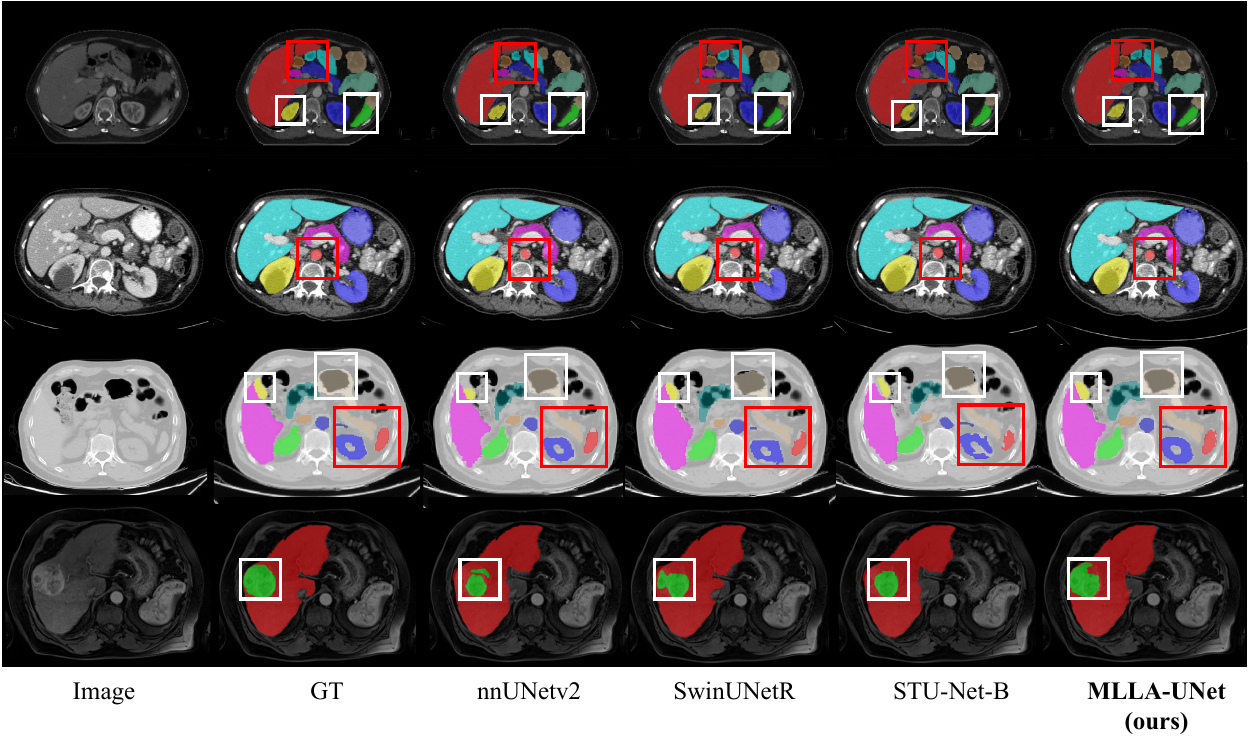}
    \caption{Visualization of segmentation results from various methods. The first three rows depict WORD, BTCV and AMOS with CT images, the forth row showcases ATLAS2023 with CE-MRI images. The six columns from left to right correspond to the original image, the ground truth (GT), the nnUNetv2 results, the SwinUNetR results, the STU-Net-B results, and our MLLA-UNet results. Red and white boxes highlight challenging regions where our method demonstrates superior performance in preserving anatomical details and boundary accuracy compared to other methods.}
    \label{fig:vis}
\end{figure}

As visually demonstrated in Figure~\ref{fig:vis}, MLLA-UNet produces more accurate and consistent segmentation results across different abdominal CT scans compared to other state-of-the-art methods like nnUNetv2~\cite{nnunetv2}, SwinUNetR~\cite{swinunetr}, and STU-Net-B~\cite{huang2023stunet}, particularly in preserving fine anatomical details and boundary definitions.
As shown in Table \ref{tab:overlall}, MLLA-UNet demonstrates consistent SOTA performance across diverse medical imaging datasets, achieving an average DSC of 88.32\%, significantly outperforming the second-best method SwinUNetR~\cite{swinunetr} (86.34\%). 
Our method excels on challenging datasets, with exceptionally high DSC scores on FLARE22~\cite{ma2024umamba} (90.15\%) and AMOS CT (90.05\%), substantially surpassing advanced models such as STU-Net-B~\cite{huang2023stunet} and SwinUNetR~\cite{swinunetr}. For AMOS MR, MLLA-UNet attains 87.29\% DSC, representing a $+$3.94\% improvement over SwinUNetR (83.35\%).
Performance gains are evident across other datasets as well. 
On ALTAS22, MLLA-UNet demonstrates a substantial improvement with 83.09\%, outperforming SwinUNetR~\cite{swinunetr} (79.15\%) by $+$3.94\%. 
For BTCV, our model achieves 85.28\%, exceeding UNetR's~\cite{unetr} performance (84.73\%, $+$0.55\%). 
On the ACDC dataset, MLLA-UNet reaches 93.28\%, surpassing MSVM-UNet~\cite{chen2024msvmunet} (92.58\%, $+$0.70\%). 
In comparison, conventional approaches like nnUNetV2~\cite{nnunetv2} and TransUNet~\cite{chen2021transunet_tran} achieve less competitive overall DSCs of 83.95\% ($+$4.37\%) and 80.01\% ($+$8.31\%) respectively, while earlier methods like nnUNetV1~\cite{nnunet} show notably lower performance (81.52\%, $+$6.80\%). 
The results across diverse imaging modalities and anatomical structures underscore MLLA-UNet's robust and superior segmentation capabilities for medical imaging applications.

\subsection{Results of BTCV multi-organ dataset
}

\begin{table}
\centering
\caption{BTCV Performance - Comparison of methods for segmenting different organs. \textbf{Bold} values indicate the best performance, and {\ul underlined} values indicate the second-best performance.}
\label{tab:btcv_overall}
\resizebox{\textwidth}{!}{%
\begin{tblr}{
  width = \linewidth,
  colspec = {lcccccccccc},
  vline{2,10} = {-}{},
  hline{1-2,15-16} = {-}{},
}
Methods & Aorta & Gallbladder & Left Kidney & Right Kidney & Liver & Pancreas & Spleen & Stomach & DSC (\%) & HD95 (mm)\\
UNet~\cite{ronneberger2015u_cnn_unet} & 85.66 & 53.24 & 81.13 & 71.60 & 92.69 & 56.81 & 87.46 & 69.93 & 74.82 & 54.59\\
Att-UNet~\cite{Attunet} & 82.61 & 61.94 & 76.07 & 70.42 & 87.54 & 46.70 & 80.67 & 67.66 & 71.70 & 34.47\\
TransUNet~\cite{chen2021transunet_tran} & 86.71 & 58.97 & 83.33 & 77.95 & 94.13 & 53.60 & 84.00 & 75.38 & 76.76 & 44.31\\
MISSFormer~\cite{huang2021missformer} & 86.99 & 68.65 & 85.21 & 82.00 & 94.41 & 65.67 & 91.92 & 80.81 & 81.96 & 18.20\\
Swin-UNet~\cite{swinunet} & 85.47 & 66.53 & 83.28 & 79.61 & 94.29 & 56.58 & 90.66 & 76.60 & 79.13 & 21.55\\
PVT-CASCADE~\cite{PVTtrans-CASCADE} & 83.01 & 70.59 & 82.23 & 80.37 & 94.08 & 64.43 & 90.10 & 83.69 & 81.06 & 20.23\\
Trans-CASCADE~\cite{PVTtrans-CASCADE} & 86.63 & 68.48 & 87.66 & 84.56 & 94.43 & 65.33 & 90.79 & 83.52 & 82.68 & 17.34\\
2D D-LKA Net~\cite{2DD-LKANet} & 88.34 & 73.79 & 88.38 & \textbf{84.92} & 94.88 & 67.71 & 91.22 & 84.94 & 84.27 & 20.04\\
MERIT-GCASCADE~\cite{MERIT-GCASCADE} & 88.05 & 74.81 & 88.01 & 84.83 & 95.38 & 69.73 & 91.92 & 83.63 & 84.54 & \textbf{10.38}\\
PVT-EMCAD-B2~\cite{rahman2024emcadPVT-EMCAD-B2} & 88.14 & 68.87 & 88.08 & 84.10 & 95.26 & \textbf{68.51} & 92.17 & 83.92 & 83.63 & 15.68\\
VM-UNet~\cite{ruan2024vmunet} & 87.00 & 69.37 & 85.52 & 82.25 & 94.10 & 65.77 & 91.54 & 83.51 & 82.38 & 16.22\\
Swin-UMamba~\cite{liu2024swinumamba} & 86.32 & 70.77 & 83.66 & 81.60 & 95.23 & 69.36 & 89.95 & 81.14 & 82.26 & 19.51\\
MSVM-UNet~\cite{chen2024msvmunet} & \uline{88.73} & \uline{74.90} & \uline{85.62} & 84.47 & \textbf{95.74} & \textbf{71.53} & \uline{92.52} & \uline{86.51} & \uline{85.00} & 14.75\\
\textbf{MLLA-UNet (Ours)} & \textbf{88.85} & \textbf{77.10} & \textbf{89.27} & \uline{84.51} & \uline{95.53} & 67.04 & \textbf{92.53} & \textbf{87.38} & \textbf{85.28} & \uline{12.96}
\end{tblr}
}
\end{table}

Table \ref{tab:btcv_overall} compares the performance of various segmentation methods on the BTCV multi-organ dataset. 
MLLA-UNet achieves the highest overall DSC of 85.28\% and the second-lowest HD95 of 12.96 mm, while MERIT-GCASCADE~\cite{MERIT-GCASCADE} obtains the lowest HD95 of 10.38 mm with a DSC of 84.54\%. 
For individual organ segmentation, MLLA-UNet achieves the highest DSC scores on multiple organs: 88.85\% for the Aorta, 77.10\% for the Gallbladder, 89.27\% for the Left Kidney, 92.53\% for the Spleen, and 87.38\% for the Stomach. 
The model obtains competitive scores of 84.51\% for the Right Kidney and 95.53\% for the Liver, ranking second in these categories. 
For Pancreas segmentation, MSVM-UNet~\cite{chen2024msvmunet} achieves the highest DSC of 71.53\%, while MLLA-UNet scores are 67.04\%. 
Compared to earlier approaches, MLLA-UNet demonstrates substantial improvements over UNet~\cite{ronneberger2015u_cnn_unet} (74.82\% DSC) and Att-UNet~\cite{Attunet} (71.70\% DSC). 
Recent methods like 2D D-LKA Net~\cite{2DD-LKANet} and PVT-EMCAD-B2~\cite{rahman2024emcadPVT-EMCAD-B2} achieve DSCs of 84.27\% and 83.63\% respectively, while MSVM-UNet obtains the second-highest overall DSC of 85.00\%.

%% file: sec/5_Discussion.tex
\section{Discussion}

\subsection{Analysis of the contribution of each architectural component}% from Table\ref{tab:overlall} and \ref{tab:btcv_overall}}
% ---
The experimental results across the diverse medical imaging datasets demonstrate MLLA-UNet's effectiveness, particularly in complex multi-organ segmentation tasks under standardized preprocessing conditions following the nnUNet pipeline. 
The superior performance can be analyzed through several critical architectural design choices concerning the experimental conditions. 
At the core of MLLA-UNet's success is its position encoding strategy, which is particularly effective in handling varied anatomical structures under different imaging modalities, as demonstrated across both CT and MRI datasets. 
The combination of LePE, CPE, and RoPE enables comprehensive spatial relationship modeling, as evidenced by the strong performance on organs with complex boundaries. 
For instance, in the BTCV dataset, with its challenging 13-organ segmentation task, MLLA-UNet achieves the highest DSC scores on the Aorta (88.85\%) and Left Kidney (89.27\%), where accurate boundary delineation is crucial. 
The HD95 metric of 12.96 mm further validates the effectiveness of our position encoding strategy in maintaining precise boundary predictions.
% ---

Complementing this encoding mechanism, the progressive channel scaling approach in EDSM ($C_i = C_0 \cdot 2^i$) and proposed EUSM ($D_i = D_L \cdot 2^{-(L-i)}$) demonstrates its value in multi-scale feature processing, particularly under the implemented data augmentation scheme including random scaling and rotation.
This is especially evident in the AMOS CT results (90.05\% DSC), where the ability to handle varying tissue contrasts and organ sizes is crucial. 
The gradual channel expansion in EDSM enables rich feature extraction at different scales. In contrast, EUSM's systematic channel reduction maintains essential information during upsampling, as reflected in the consistent performance across the diverse dataset collection comprising FLARE22, AMOS22, ATLAS23, WORD, BTCV, and ACDC. 
However, despite the implemented data augmentation strategies, the relatively lower performance in pancreas segmentation (67.04\% DSC) suggests potential limitations in handling highly variable and small anatomical structures. 
This might be attributed to the trade-off between spatial resolution reduction in EDSM and the preservation of fine-grained features necessary for small organ segmentation. 
Future improvements could focus on adaptive channel scaling strategies that better preserve detailed features for challenging anatomical structures while maintaining the current advantages in global context modeling, potentially through modifications to the learning rate schedule or batch size configurations.

\subsection{Ablation Study}

\subsubsection*{Model scaling and backbone comparison}
\begin{table}[!htbp]
\centering
\caption{Performance comparison of various medical image segmentation methods across multiple datasets. The table shows the method type, model parameters (in millions), FLOPs (in billions), and Dice scores (\%) for each dataset. The best results for each dataset are highlighted in \textbf{bold}.}
\label{tab:ablation}
\resizebox{\textwidth}{!}{%
\begin{tblr}{
  colspec = {lccccccccccc},
  vline{2-4} = {-}{},
  hline{1-2,8} = {-}{},
}
Methods & Type & Params & GFLOPs & WORD & FLARE22 & AMOSCT & ALTAS23 & BTCV & ACDC & AMOSMR & Avg\\
ConvNeXtv2~\cite{convnextv2} & CNN & 48.51M & 23.82 & 88.37 & 87.24 & 87.15 & 79.14 & 81.25 & 89.15 & 86.36 & 85.52\\
Swin~\cite{liu2021swin} & Transformer & 33.72M & 17.84 & 84.15 & 85.27 & 85.31 & 70.15 & 78.51 & 87.20 & 84.16 & 82.11\\
VSS~\cite{liu2024vmamba_94} & Mamba & 44.27M & 11.55 & 87.85 & 86.13 & 86.77 & 78.91 & 81.05 & 88.81 & 85.12 & 84.95\\
MLLA\textsubscript{Tiny} & MLLA & 34.14M & 14.66 & \textbf{89.10} & \textbf{90.15} & \textbf{90.05} & 83.09 & \textbf{85.28} & \textbf{93.28} & 87.29 & \textbf{88.32}\\
MLLA\textsubscript{Small} & MLLA & 64.52M & 26.30 & 84.59 & 89.33 & 87.45 & \textbf{84.21} & 83.41 & 91.32 & \textbf{88.33} & 86.95\\
MLLA\textsubscript{Base} & MLLA & 144.5M & 58.56 & 83.18 & 89.25 & 86.31 & 79.31 & 82.63 & 90.19 & 85.34 & 85.17
% EncOnly & MLLA & 34.14M & 14.66G & 84.82 & 89.1 & 88.51 & 80.19 & 83.21 & 91.05 & 86.25 & 86.16
\end{tblr}
}
\end{table}

% ---

Table \ref{tab:ablation} presents an ablation study focused on analyzing the effectiveness of our MLLA blocks and their configuration within the overall architecture, as shown in Figure~\ref{fig:model}. 

We conduct experiments comparing different variants: replacing MLLA blocks with traditional CNN (ConvNeXtv2~\cite{convnextv2}), transformer (Swin~\cite{liu2021swin}), and  Mamba-style sequence modeling (VSS~\cite{liu2024vmamba_94}) components while maintaining the same U-shaped architecture. 
Additionally, we investigate the impact of model capacity by scaling the MLLA architecture to different sizes (Tiny, Small, and Base).

Comparing MLLA\textsubscript{Tiny} with traditional CNN-based methods like ConvNeXtv2, we observe a significant performance improvement of 2.8 percentage points in average Dice score (88.32\% vs. 85.52\%). 
This improvement can be attributed to our linear attention mechanism maintaining $\mathcal{O}(N)$ complexity while effectively modeling long-range dependencies, combined with our comprehensive position encoding strategy. 
The MLLA\textsubscript{Tiny} model also outperforms transformer-based (Swin) and SSM-based (VSS) architectures, with improvements of 6.21 and 3.37 percentage points, respectively. 
These gains demonstrate the effectiveness of the MLLA triple position encoding approach: the local spatial information captured by LePE through depth-wise convolution, the input-dependent contextual information from CPE, and the global positional awareness provided by RoPE collectively enable more effective feature representation for medical image segmentation.

% ---
Interestingly, we observe that simply scaling up the model size by increasing the embedding dimensions and the number of MLLA blocks at each layer (MLLA\textsubscript{Small} and MLLA\textsubscript{Base}) does not necessarily lead to better performance.
While MLLA\textsubscript{Small} shows some improvement over MLLA\textsubscript{Tiny} in specific datasets (e.g., AMOSMR), the overall average performance decreases.
Recent empirical evidence from Gao et al. \cite{gao2024training} corroborates our observations - merely expanding model architectures yields diminishing returns when constrained by limited training data quantity and variety.
This observation presents an interesting departure from conventional wisdom regarding neural network scaling \cite{dosovitskiy2020image_vit,kaplan2020scaling,liu2021swin,he2016deep}, where the relationship between model capacity and dataset characteristics plays a crucial role in preventing overfitting issues.
Table \ref{tab:model_architecture} describes the detailed scale-up method.
We carefully analyze the reasons for this performance decline and propose solutions in the following Section~\ref{sec:discussion}.
% ---

\subsubsection*{Ablation study on up-sampling and down-sampling strategies}

\begin{table}[!t]
\small
\centering
\caption{Evaluation of different up-sampling and down-sampling operations on the WORD dataset, reporting DSC (\%), HD95 (mm), GFLOPs of different modules, and the corresponding number of parameters (\#Param. in \textit{thousand}). The best results are highlighted in \textbf{bold}, and the second-best results are in \uline{underlined}.}
\label{tab:ablationdownup}
\begin{tblr}{
  colspec = {lcccc},
  vline{2,4} = {-}{},
  hline{1-2,6-7,9} = {-}{},
}
Up-sampling Operations & \#GFLOPs & \#Param. (K) & DSC (\%) & HD95 (mm)\\
Patch Expand~\cite{swinunet} & \textbf{1.27} & 18.5 & 85.31 & 19.48\\
LKPE~\cite{chen2024msvmunet} & 12.58 & 21.0 & \uline{88.39} & \uline{12.21}\\
EUCB~\cite{rahman2024emcadPVT-EMCAD-B2} & 3.60 & \uline{15.5} & 86.81 & 18.20\\
\textbf{EUSM (Ours)} & \uline{1.77} & \textbf{13.7} & \textbf{89.10} & \textbf{9.37}\\
Down-sampling Operations & \#GFLOPs & \#Param. (K) & DSC (\%) & HD95 (mm)\\
Patch Merge~\cite{swinunet} & \textbf{1.21} & 74.5 & 88.03 & 14.37\\
\textbf{EDSM (Ours)} & 1.66 & \textbf{52.4} & \textbf{89.10} & \textbf{9.37}
\end{tblr}
\end{table}

%---
In enhancing encoder-decoder architectures for medical image segmentation, we conducted a comprehensive ablation study on the WORD dataset, focusing on the efficacy of various up-sampling and down-sampling operations within our MLLA-UNet model. 
The results are detailed in Table \ref{tab:ablationdownup}. This study evaluates critical performance metrics, including DSC, HD95, the computational demand in GFLOPs, and the total number of parameters.
%---

% ---
Our proposed EUSM outperformed the alternatives for up-sampling operations, achieving the highest DSC at 89.1\%. % and the lowest HD95 at 9.37 mm. 
This indicates superior segmentation accuracy and enhanced boundary delineation. 
Despite its high accuracy, our method also maintained remarkable computational efficiency, using only 1.77 GFLOPs and the smallest model size of 13.7K parameters. 
In contrast, while being the most computationally efficient at 1.27 GFLOPs, the Patch Expand strategy significantly lagged behind in segmentation performance, with a DSC of 85.31\% % and HD95 of 19.48 mm. 
The LKPE strategy showed a good balance with a DSC of 88.39\% and an HD95 of 12.21 mm but was considerably more computationally demanding at 12.58 GFLOPs. 
The EUCB method provided a moderate performance across the metrics but was still outshone by our proposed method in every aspect except GFLOPs.
% ---

% ---
In terms of downsampling operations, our EDSM also excelled, achieving a DSC of 89.1\%, substantially higher than the traditional Patch Merge strategy's 88.03\% and a significantly reduced HD95 of 9.37 mm compared to 14.37 mm. 
Although our method required slightly more computational power (1.66 GFLOPs versus 1.21 GFLOPs), it reduced the model's parameter count from 74.5K to 52.4K, enhancing model efficiency and compactness. 
Overall, our ablation study demonstrates that our proposed up-sampling and down-sampling strategies in the MLLA-UNet architecture optimize segmentation accuracy and boundary precision, improve computational efficiency, and reduce model size. These results validate our architectural innovations and underscore their potential in advancing the field of medical image segmentation.
% ---

\subsection{Scalability of the proposed MLLA-UNet}
\subsubsection*{Scaling model and dataset simultaneously for improved performance}
\label{sec:discussion}
% ---
In addressing the challenges of model scaling as illustrated in Table~\ref{tab:ablation}, we adopted a strategy inspired by Huang et al. ~\cite{huang2023eval}, where both the model size and the dataset were expanded concurrently. 
The results in Table \ref{tab:aeval} evaluate the performance across multiple datasets focusing on shared organ categories. 
Notably, when trained with this expanded dataset, the larger MLLA\textsubscript{Base} model achieved the highest performance, recording an average Dice score of 90.28\% %. 
This outcome underscores the efficacy of combining increased model capacity with diverse training datasets to counteract overfitting and enhance generalization capabilities effectively.
% ---

% ---
The performance improvements are particularly striking for anatomically complex organs like the pancreas and gallbladder, where the Dice scores reached 88.7\% and 80.21\%, respectively. 
These results suggest that the enhanced capacity of the MLLA\textsubscript{Base} model, when paired with a varied dataset, can more accurately represent intricate anatomical features and relationships. 
In conclusion, the ablation studies underscore the prowess of the MLLA architecture, especially the MLLA\textsubscript{Tiny} variant, in delivering top-tier performance while maintaining competitive computational efficiency. 
Moreover, our approach of scaling both the model size and dataset diversity has proven to be a successful strategy for achieving superior performance, as demonstrated by the outstanding results of the MLLA\textsubscript{Base} model in our expanded evaluation framework.
% ---

\begin{table}[!htbp]
\small
\centering
\caption{Comparison of performance for different training methods using \cite{huang2023eval}. The methods were evaluated by expanding both model size and dataset size, and training was conducted using the same shared organ categories across all datasets.}
\label{tab:aeval}
\resizebox{\textwidth}{!}{%
\begin{tblr}{
  width = \linewidth,
  colspec = {lccccccccc},
  vline{2} = {-}{},
  hline{1-2,5} = {-}{},
}
Methods & Liver & Right Kidney & Spleen & Pancreas & Gallbladder & Esophagus & Stomach & Left Kidney & Avg\\
MLLA\textsubscript{Tiny} & 96.80 & 94.57 & 93.72 & 87.41 & 78.83 & 79.61 & 93.21 & 92.41 & 89.57\\
MLLA\textsubscript{Small} & 97.12 & \textbf{95.21} & 93.90 & 87.88 & 79.69 & 79.83 & 94.28 & 91.89 & 89.98\\
MLLA\textsubscript{Base} & \textbf{96.50} & 95.09 & \textbf{94.03} & \textbf{88.70} & \textbf{80.21} & \textbf{80.18} & \textbf{94.36} & \textbf{93.14} & \textbf{90.28}
\end{tblr}
}
\end{table}

\begin{table}[!t]
    \caption{Architectures of MLLA-UNet models.}
    \label{tab:model_architecture}
    \vskip 0.1in
    \centering
    \footnotesize
    \setlength{\tabcolsep}{2mm}{
    \renewcommand\arraystretch{1.5}
    \begin{tabular}{c|c|c|c|c|c}
    \toprule
    \textbf{stage} & \textbf{output} & \textbf{scale} & \textbf{MLLA-UNet\textsubscript{Tiny}} & \textbf{MLLA-UNet\textsubscript{Small}} & \textbf{MLLA-UNet\textsubscript{Base}} \\
    \midrule
    input & $224\times224$ & 1 & \multicolumn{3}{c}{input image} \\
    \midrule
    \multirow{3}*{res1} & \multirow{3}*{$56\times 56$} & \multirow{3}*{1/4 ↓} & stem, 64 & stem, 64 & stem, 96 \\
    \cline{4-6}
    && & $\left[\!\!\! \begin{array}{c} {\rm dim} \ 64 \\ {\rm head} \ 2\end{array} \!\!\! \right ] \!\!\times\! 2$ & $\left[\!\!\! \begin{array}{c} {\rm dim} \ 64 \\ {\rm head} \ 2\end{array} \!\!\! \right ] \!\!\times\! 3$ & $\left[\!\!\! \begin{array}{c} {\rm dim} \ 96 \\ {\rm head} \ 3\end{array} \!\!\! \right ] \!\!\times\! 3$ \\
    \midrule
    \multirow{3}*{res2} & \multirow{3}*{$28\times 28$} & \multirow{3}*{1/8 ↓} & EDSM, 128 & EDSM, 128 & EDSM, 192 \\
    \cline{4-6}
    && & $\left[\!\!\! \begin{array}{c} {\rm dim} \ 128 \\ {\rm head} \ 4\end{array} \!\!\! \right ] \!\!\times\! 4$ & $\left[\!\!\! \begin{array}{c} {\rm dim} \ 128 \\ {\rm head} \ 4\end{array} \!\!\! \right ] \!\!\times\! 6$ & $\left[\!\!\! \begin{array}{c} {\rm dim} \ 192 \\ {\rm head} \ 6\end{array} \!\!\! \right ] \!\!\times\! 6$ \\
    \midrule
    \multirow{3}*{res3} & \multirow{3}*{$14\times 14$} & \multirow{3}*{1/16 ↓} & EDSM, 256 & EDSM, 256 & EDSM, 384 \\
    \cline{4-6}
    && & $\left[\!\!\! \begin{array}{c} {\rm dim} \ 256 \\ {\rm head} \ 8\end{array} \!\!\! \right ] \!\!\times\! 8$ & $\left[\!\!\! \begin{array}{c} {\rm dim} \ 256 \\ {\rm head} \ 8\end{array} \!\!\! \right ] \!\!\times\! 21$ & $\left[\!\!\! \begin{array}{c} {\rm dim} \ 384 \\ {\rm head} \ 12\end{array} \!\!\! \right ] \!\!\times\! 21$ \\
    \midrule
    \multirow{4}*{res4} & \multirow{4}*{$7\times 7$} & \multirow{4}*{1/32 ↓} & EDSM, 512 & EDSM, 512 & EDSM, 768 \\
    \cline{4-6}
    && & $\left[\!\!\! \begin{array}{c} {\rm dim} \ 512 \\ {\rm head} \ 16\end{array} \!\!\! \right ] \!\!\times\! 4$ & $\left[\!\!\! \begin{array}{c} {\rm dim} \ 512 \\ {\rm head} \ 16\end{array} \!\!\! \right ] \!\!\times\! 6$ & $\left[\!\!\! \begin{array}{c} {\rm dim} \ 768 \\ {\rm head} \ 24\end{array} \!\!\! \right ] \!\!\times\! 6$ \\
    \cline{4-6}
    && & EUSM, 512 & EUSM, 512 & EUSM, 768 \\
    \midrule
    \multirow{3}*{res5} & \multirow{3}*{$14\times 14$} & \multirow{3}*{1/16 ↑} & $\left[\!\!\! \begin{array}{c} {\rm dim} \ 256 \\ {\rm head} \ 8\end{array} \!\!\! \right ] \!\!\times\! 8$ & $\left[\!\!\! \begin{array}{c} {\rm dim} \ 256 \\ {\rm head} \ 8\end{array} \!\!\! \right ] \!\!\times\! 21$ & $\left[\!\!\! \begin{array}{c} {\rm dim} \ 384 \\ {\rm head} \ 12\end{array} \!\!\! \right ] \!\!\times\! 21$ \\
    \cline{4-6}
    && & EUSM, 256 & EUSM, 256 & EUSM, 384 \\
    \midrule
    \multirow{3}*{res6} & \multirow{3}*{$28\times 28$} & \multirow{3}*{1/8 ↑} & $\left[\!\!\! \begin{array}{c} {\rm dim} \ 128 \\ {\rm head} \ 4\end{array} \!\!\! \right ] \!\!\times\! 4$ & $\left[\!\!\! \begin{array}{c} {\rm dim} \ 128 \\ {\rm head} \ 4\end{array} \!\!\! \right ] \!\!\times\! 6$ & $\left[\!\!\! \begin{array}{c} {\rm dim} \ 192 \\ {\rm head} \ 6\end{array} \!\!\! \right ] \!\!\times\! 6$ \\
    \cline{4-6}
    && & EUSM, 128 & EUSM, 128 & EUSM, 192 \\
    \midrule
    \multirow{3}*{res7} & \multirow{3}*{$56\times 56$} & \multirow{3}*{1/4 ↑} & $\left[\!\!\! \begin{array}{c} {\rm dim} \ 64 \\ {\rm head} \ 2\end{array} \!\!\! \right ] \!\!\times\! 2$ & $\left[\!\!\! \begin{array}{c} {\rm dim} \ 64 \\ {\rm head} \ 2\end{array} \!\!\! \right ] \!\!\times\! 3$ & $\left[\!\!\! \begin{array}{c} {\rm dim} \ 96 \\ {\rm head} \ 3\end{array} \!\!\! \right ] \!\!\times\! 3$ \\
    \cline{4-6}
    && & final patch expand, 64 & final patch expand, 64 & final patch expand, 96 \\
    \bottomrule
    \end{tabular}}
\end{table}

%% file: sec/6_Conclusion.tex
\section{Conclusion and Future Works}

In this paper, we introduced MLLA-UNet, a novel architecture for medical image segmentation that integrates Mamba-inspired designs and linear attention mechanisms. Our approach efficiently processes high-resolution images while accurately capturing long-range dependencies and preserving local structural information. The core innovation of MLLA-UNet lies in its hybrid architecture that combines the advantages of linear attention and State Space Models (SSMs), achieving linear computational complexity $O(n)$ while maintaining high expressiveness in feature extraction. We further enhanced the model's capabilities through an innovative symmetric sampling structure featuring Efficient DownSampling Module (EDSM) and Efficient UpSampling Module (EUSM), effectively preserving spatial information and enabling precise multi-scale feature fusion. To strengthen spatial relationship modeling, we incorporated sophisticated position encoding strategies, including Local Positional Encoding (LePE), Conditional Positional Encoding (CPE), and Rotary Position Encoding (RoPE). Extensive experiments across six challenging datasets with 24 different segmentation tasks validate our approach, with our MLLATiny variant achieving an average DSC of 88.32\% using only 34.14M parameters and 14.66G FLOPs, significantly outperforming existing state-of-the-art models. Our scalability analysis reveals essential insights into model scaling and dataset size relationships, establishing MLLA-UNet as a robust and efficient solution for complex medical image segmentation tasks, particularly well-suited for emerging clinical applications and challenging anatomical structures that have traditionally been difficult to segment accurately.

The cross-modal adaptability of our model may streamline workflows in clinical settings where multiple imaging modalities are used \cite{MartiBonmati2010}, potentially reducing interpretation time and improving patient care. Additionally, the computational efficiency of MLLA-UNet facilitates real-time or near-real-time segmentation crucial for time-sensitive applications such as image-guided interventions or emergency radiology \cite{Xu2019}. By improving segmentation accuracy and efficiency across various medical imaging modalities, MLLA-UNet supports better treatment planning, enhances patient outcomes, and contributes to the advancement of personalized medicine \cite{Norgeot2019}, aligning with broader trends in leveraging artificial intelligence to enhance radiological practice and patient care \cite{Hosny2018}. As we continue to refine and expand MLLA-UNet, we anticipate that this innovative approach will play a crucial role in the ongoing revolution in healthcare technology, with future research directions including extending the architecture to 3D segmentation, adapting it for lightweight video streams, incorporating multi-modal fusion techniques, exploring self-supervised learning approaches, integrating explainable AI techniques, and developing adaptive architectures for resource-constrained environments.